\documentstyle[12pt,aasms4]{article}
\received{}
\accepted{}
\journalid{}
\articleid{}
\lefthead{CHARTAS et al.}
\righthead{X-RAY DETECTION OF Q0957+561 LENS}

\begin{document}


\def\beginrefer{\section*{REFERENCES}%
\begin{quotation}\mbox{}\par}
\def\refer#1\par{{\setlength{\parindent}{-\leftmargin}\indent#1\par}}
\def\endrefer{\end{quotation}}

\title{X-ray Detection of the Primary Lens Galaxy Cluster of the 
Gravitational Lens System Q0957+561}

\author{G. Chartas\altaffilmark{1}, D. Chuss\altaffilmark{2}, W. Forman\altaffilmark{3}, 
C. Jones\altaffilmark{3}, and I. Shapiro\altaffilmark{3}}

\altaffiltext{1}{Astronomy and Astrophysics Department, The Pennsylvania
State University, 525 Davey Laboratory, University Park, PA 16802.}
\authoremail{gchartas@astro.psu.edu}

\altaffiltext{2}{Department of Physics and Astronomy, Northwestern University,
2145 Sheridan Road, Evanston, Illinois 60208-3112.} 

\altaffiltext{3}{Smithsonian Astrophysical Observatory,
Harvard-Smithsonian Center for Astrophysics, 60 Garden Street, Cambridge, MA 02138.}

\begin{abstract}
Analysis of several recent ROSAT HRI observations of the gravitationally lensed 
system Q0957+561 has led to the detection at the 3$\sigma$ level of the 
cluster lens containing the primary galaxy G1.   
The total mass was estimated by applying the equation
of hydrostatic equilibrium to the detected hot intracluster gas 
for a range of cluster core radii, cluster sizes
and for different values of the Hubble constant.

X-ray estimates of the lensing cluster mass provide a 
means to determine the cluster contribution to the deflection 
of rays originating from the quasar Q0957+561. The present mass estimates 
were used to evaluate the convergence parameter $\kappa$,
the ratio of the local surface mass density of the cluster to the 
critical surface mass density for lensing. The convergence parameter, $\kappa$,
calculated in the vicinity of the lensed images, was found to range between 0.07 and
0.21, depending on the assumed cluster core radius and cluster extent. 
This range of uncertainty in $\kappa$ does not include 
possible systematic errors arising from the estimation of the cluster temperature
through the use of the cluster luminosity-temperature relation
and the assumption of spherical symmetry of the cluster gas. 
Applying this range of values of $\kappa$ to the lensing model of
Grogin \& Narayan (1996) for Q0957+561 but not accounting for uncertainties
in that model yields a range of values for the Hubble constant: $ 67 <$
H$_{0}$ $<$ 82 $\,$ km $\,$ s$^{-1}$ $\,$ Mpc$^{-1}$, for a time delay of 1.1 years.

\end{abstract}

\section{INTRODUCTION}

Refsdal (1964a 1964b 1966) outlined a
method for determining a global value for the Hubble constant by measuring 
the light travel delay between different images of a gravitationally lensed system.
Together with a model that describes the gravitational potential of
the lens, the time delay defines the geometrical dimensions
of the lens system and therefore may lead to the determination of the
Hubble constant.

The criteria that make a gravitational lens system (GLS) suitable for
measuring the Hubble constant can be summarized as follows:
(a) The sky positions and redshifts of the lens and of the images of 
the lensed object should be measurable accurately; (b) The mass 
distribution of the lens should be subject to accurate and reliable 
estimation. When the lens has multiple components, as for Q0957+561,
this estimation becomes especially difficult and benefits from
many different types of observations and also from many and
extended images, such as rings. For example, if the types of observations 
include radio and optical emission lines, which are less sensitive
to microlensing, flux ratio measurements are more useful in constraining
the lens model; and (c) The differences in propagation time from the lensed
object to us through (at least two) different images (``time delay'')
should be measurable with high accuracy. The source, i.e., the lensed
object, must therefore be variable on time scales far shorter
than the time delays, to allow accurate estimates to be made 
of the latter.

Several potential candidates for determining $H_{0}$
are the quasar Q0957+561 with two images and a measured time delay
of 1.1 years (Pelt et al. 1996; Haarsma et al. 1997; Kundi\'{c} et al. 1997), 
the quadruple image system PG 1115+080 with a measured time delay
between components B and C of about 25 days (Schechter et al. 1997; 
Bar-Kana 1997), and B0218+357 with a time delay of
12 $\pm$ 3 days (Geiger \& Schneider 1996; Corbett et al. 1996).
Combining measured time delays with a
detailed lensing model, Refsdal's suggestion has been applied to
Q0957+561 (Falco, Gorenstein, \& Shapiro 1991; Grogin \& Narayan 1996, hearafter GN) 
and PG1115+080 (Keeton \& Kochanek 1997).

In this paper we focus on a study of the properties of 
Q0957+561 (Walsh, Carswell \& Weyman 1979), one of the most extensively studied
lensing systems which has been monitored since its discovery in both the optical and
radio (see, for example, Schild \& Thomson 1995; Vanderriest et al. 1989; Haarsma et al. 1997). 
The primary lensing components are an elliptical cD galaxy at a redshift of 0.355,
usually referred to as the G1 galaxy, a cluster of galaxies containing
the G1 galaxy, and a group of galaxies at a redshift of 0.5.  The most 
thorough optical spectroscopic and photometric study to date of the field
around Q0957+561 has been presented by Angonin-Willaime, Soucail, \& Vanderriest (1994).

One of the major concerns in modeling Q0957+561 is a mass sheet degeneracy
which arises from the presence of the cluster of  galaxies at z=0.36.
In particular if one were to modify the assumed radial mass density
profile $\kappa(\theta)$ of the lens to ${\lambda}{\kappa(\theta)}$ + (1 - $\lambda$),
where 1 - $\lambda$ represents  a constant surface mass density sheet term, then the new mass
distribution would also satisfy the observational constraints.
However the resulting value for $H_{0}$ would be scaled by the constant 1 - $\lambda$.
Since $\lambda$ must always be positive one may always determine an upper
limit on the Hubble constant for $\lambda$ = 0 (Falco et al. 1991).

As also noted by Falco et al. (1991) and later GN, this degeneracy may 
be broken if one were to measure the velocity dispersion or total 
mass distribution of either the cluster of galaxies
or the principal lensing galaxy G1.
Recently Falco et al. (1997) have obtained a velocity dispersion for 
the central lens galaxy G1 of $\sigma_{v} = 266 \pm 12$ km s$^{-1}$
or $\sigma_{v} = 279 \pm 12$ km s$^{-1}$ depending on the interpretation of 
the spectroscopic slit data.\\ 
  
In this paper we focus on improving the cluster model
and we present an estimate of the total mass
of this lensing cluster of galaxies through X-ray measurements of thermal emission
from the intracluster gas. In section 2 we describe the spatial 
analysis of the X-ray data of Q0957+561
obtained from several ROSAT HRI observations and we present an estimate for the
convergence parameter $\kappa(\theta)$ of the cluster.
Section 3 describes the spectroscopic analysis of ROSAT position
sensitive proportional counter (PSPC) and
ASCA Gas Imaging Spectrometer (GIS) X-ray observations at different
epochs in order to investigate the variability of different spectral components,
and presents various scenarios to explain the observed variability in the X-ray flux.
Finally section 4 presents our conclusions from
the analysis of the X-ray observations of the  Q0957+561
GL system. In particular, we discuss how our results allow
bounds to be placed on the Hubble constant, albeit with these bounds
still subject to systematic errors that we cannot yet evaluate accurately.

\section{SPATIAL ANALYSIS: A 3$\sigma$ DETECTION OF THE CLUSTER THAT CONTAINS THE G1 GALAXY}
The limits on the mass and luminosity of the cluster of galaxies that contains the
primary lensing cD elliptical galaxy G1 presented in Chartas et al. (1995) were 
based on the ROSAT HRI observation of Q0957+561 from 1992 May 1. 
We now estimate the mass and luminosity
of this cluster utilizing additional 
data obtained from the three subsequent ROSAT HRI observations 
in 1992 October 20, 1992 October 22, and 1994 April 11 
(see Table 1).
The October 20 ROSAT HRI observation is centered on the galaxy NGC 3079 
with the quasar Q0957+561 located 12 arcmin off-axis. Because of the dependence of the 
properties of the point spread function (PSF) with off-axis angle, we performed 
the analysis for this off-axis 
observation separately and combined the results with those obtained
from the on-axis HRI observations of Q0957+561. To look for extended cluster 
emission centered on the galaxy G1, we first merged the  
on-axis observations of 1994 April, 1992 October, and 1991 May.
The image files containing the data for these observations were shifted 
appropriately such that the center of Q0957+561 image B from
each observation was co-aligned.
The shifts were relatively small, of the order of a few arcsecs.
Figure 1 shows the resulting merged image of the three HRI observations.
X-ray emission surrounding the lensed images can be seen. 
To determine the significance of this extended emission
and whether it originates from hot cluster gas emission or from
the PSF wings of the lensed images, we searched  for a statistically 
significant excess in detected counts
originating from cluster emission around the G1 galaxy.
In particular we determined the number of counts within annuli 
of inner radii ranging from 15$''$ to 100$''$ in steps of 5$''$ 
and outer radius of 120$''$  
centered on Q0957+561 for the merged ROSAT HRI observations and separately for
the off-axis observation.
 The background was taken from an annulus of inner and outer 
radii of 340$''$ and 460$''$. 
For inner radii less than about 25$''$ the contribution from the 
PSF's of images A and B of Q0957+561 is significant whereas  
for inner radii greater than about 60$''$ the signal to noise drops considerably 
leading to a large uncertainty in the determined source counts, as seen in Figure 2.
We modeled two point sources at quasar image locations A and B.
Each modeled source PSF was normalized such that 
the number of counts within a radius of 3$''$ from the center of the 
modeled source was equal to the number of counts within a radius of 
3$''$ of the observed source.
The net counts $N_{cl}{(r_{12})}$ detected by the ROSAT HRI originating 
from cluster emission within an annulus of inner radius $r_{1}$ and outer 
radius $r_{2}$ were determined by subtracting background and point source contributions 
and is given by the expression,

\begin{equation}
{N_{cl}{(r_{12})}} = {N_{tot}}(r_{12}) - {N_{bkg}}(r_{12}) - {N_{A}}(r_{12})  - {N_{B}}(r_{12})
\end{equation}

\noindent
where ${N_{bkg}}(r_{12})$ are the background counts in the annular region $r_{12}$ 
and ${N_{A}}(r_{12})$, ${N_{B}}(r_{12})$  
are the counts within the region $r_{12}$ originating from the A and B quasar images
of Q0957+561, respectively.   

To compute the total cluster counts, we estimated a correction term for
the cluster counts beyond the range of the inner and outer radii of the annulus 
(as described in Chartas et al. 1995). 
The correction term depends on the adopted value for the cluster 
core radius and cluster radius. For our calculations we consider core radii ranging between
$r_{c} = 5''$ and $r_{c} = 45''$ and  cluster radii ranging between 
120$''$ and 280$''$.
For a typical cluster core radius of $r_{c} = 15''$ (0.067 Mpc, assuming 
$H_{0} = 75$ km s$^{-1}$ Mpc$^{-1}$ and $q_{0} = 0$), we find that the 
number of counts within a 20$''$ circle centered on Q0957+561 is about 60$\%$  of the 
counts found in the annulus with inner radius of 20$''$ and outer radius of  120$''$, 
whereas for $r_{c} = 35''$ we find that about 18 $\%$ of the X-ray counts originate within 20$''$.
No X-ray sources detected within the HRI field 
of view lie within the annuli used as source and background extraction regions.
For the four HRI observations of Q0957+561 with a total exposure time of 
85154sec, we estimate that a total of about 300 $\pm$ 100 counts (with the quoted error being at
the 1$\sigma$ level) originated from cluster emission and approximately 2900 counts
from the quasar images.

To convert the measured HRI counts into luminosities we have assumed a 
Raymond-Smith spectrum characterized by a temperature $T_{e}$, 
a galactic column density $N_{H}$ = $0.87 \times 10^{20} cm^{-2}$ (Stark et al. 1992), 
30$\%$ cosmic abundances (see, e.g., Henriksen 1985; Hughes et al. 1988; and Arnaud et al. 1987)
and the $L_{X}$ - $T_{e}$ correlation determined by David et al. (1993). 
The relation between cluster luminosity and temperature for a value of the Hubble constant
of $H_{0}$ = 100h km s$^{-1}$ Mpc$^{-1}$ is, $T_{e}(keV) = 10^{-12.0254}(h^{2}L_{X})^{b}$  with
$b$ = 0.29.


There is considerable scatter in the observational data used by David et al. (1993)
to produce the $L_{X}$ vs. $T_{e}$ correlation.
A recent study however by Markevitch (1998) has shown that the 
scatter in the $L_{X}$ -  $T_{e}$ relation is greatly reduced when
the emission from cooling flow regions is excluded in the analysis.
To derive the cluster mass, we assume a typical density profile
$\rho(r)$ of the hot gas (e.g., Jones \& Forman 1996) given by :

\begin{equation}
\rho_{gas}(r)={\rho_o\over{\left({1+r/r_c}\right)^{3\beta/2}}}
\end{equation}
\noindent
and apply the equation of hydrostatic equilibrium to
the intracluster hot gas, which we assume to be isothermal.
The total mass of the lens within a radius r is given by

\begin{equation}
{M_{grav}(<r)}  = { {{3 \beta k T}\over{\mu m_{p} G}} { {r} \over {[1 + ({{r_c}\over{r}})^2]}}} {\  ,}
\end{equation}
where $r_c$ is the core radius of the cluster, $\mu$$m_{p}$ is the mean molecular weight of the gas, 
and $\beta$ is the ratio of kinetic energy per unit mass in galaxies to 
kinetic energy per unit mass in the gas.

The derived cluster temperature depends on the assumed value of $H_{0}$ and 
the assumed core radius of the cluster.
Therefore, we determined the cluster mass within a radius of 1Mpc
for a range of cluster core radii and for $H_{0}$ values of 50 and
75 km s$^{-1}$ Mpc$^{-1}$. The results of this analysis are shown in Figure 3.

\normalsize

We use the estimate of the cluster mass to evaluate the convergence parameter $\kappa$,
the ratio of the local surface mass density of the cluster to the
critical surface mass density for lensing.
The convergence parameter together with the time delay $\Delta\tau_{BA}$
of the GLS are two remaining principal parameters with the largest
uncertainties  that enter in the gravitational models
recently developed by GN used to determine the Hubble constant.
GN obtain,
\begin{equation}
H_{0}  = (85^{+6}_{-7})(1-\kappa)({1.1yr}/\Delta\tau_{BA}) \,km\,s^{-1}\,Mpc^{-1}.  
\end{equation}

Recent deep HST observations of Q0957+561 by Bernstein et al. (1997) have detected several
new features and provide improved positions for the lens components. In particular the 
position of the center of G1 is found to lie at the position previously
determined by VLBI (see, e.g., Falco et al. 1991) and significantly distant from 
the position used in the GN derivation of equation 4.   
The previously published estimates for $\kappa$ are very uncertain because they
depend (strongly) on the assumption that the cluster potential may be represented
as a softened isothermal sphere. The velocity dispersion and its relation to the mass 
of the cluster, the cluster core radius, the cluster shape and the position of the 
cluster center with respect to the lensed object also need to be known. 
The present HRI observations provide a partly independent, but still 
model-dependent, estimate of the total mass of the lens and hence of the 
convergence parameter $\kappa$ through the expression

\begin{equation}
{\kappa(x) = {{\Sigma(x)}\over{\Sigma_{cr}}}} \,
\end{equation}
where ${\Sigma(x)}$ is the surface mass density of the cluster lens and 
$\Sigma_{cr}$ is the critical surface mass density (see, e.g., Blandford $\&$ Narayan 1992), 
the density the cluster must have to produce multiple images if it were the only lensing agent, and
is given by:

\begin{equation}
{\Sigma_{cr}={{c^2}\over{4\pi G}}{D_{os}\over{D_{ds}D_{od}}}}
\end{equation}  
where D$_{os}$, D$_{ds}$, and D$_{od}$ are the angular-diameter distances between source and observer,
deflector and source, and deflector and observer, respectively. We evaluated the 
critical surface mass density for the gravitationally lensed system Q0957+561 to be
$\Sigma_{cr}$ = 0.9452(0.9224)$h$ gr cm$^{-2}$, 
for $q_{0}$ = 0(0.5).

In estimating $H_{0}$ from the GLS, we need to know the cluster
contribution to the lensing of the quasar. In this treatment, 
we linearly superpose the effects 
of G1 and the cluster. We first view G1's contribution to the lensing as producing
a pair of images, A$'$ and B$'$, as shown in Figures 4 and 5.  
The contribution of the central galaxy G1 has been represented by SPLS
and FGS models (GN, Falco et al. 1991).  


For the cluster, we assume a spherically symmetric potential 
which produces an outward radial offset of the lensed components
with respect to the cluster center (see Figures 4 and 5). 
To simplify the calculations we incorporate the thin lens approximation 
in which the cluster is projected onto a plane perpendicular to the line 
of sight, on the plane of the sky.

To calculate the angle, $\vec{\alpha}$, by which a ray is deflected  
by the cluster lens, we integrated the deflection
over the entire cluster. The scaled deflection angle (Schneider, Ehlers, \& Falco 1992) is given by

\begin{equation}
{\vec{\alpha}(\vec{x})}={1\over\pi}{\int {d^2x'\,\kappa({\vec{x'}}) {{{\vec {x}-\vec{x}'}}\over{|{\vec {x}-\vec{x}'}}|^2}}}
\end{equation}
where ${\vec{x'}}$ represents the scaled location of a projected 
mass element in the lens plane with respect to the cluster center 
and ${\vec{x}}$ represents the scaled location of the impact ray 
which is deflected by $\vec{\alpha}(\vec{x})$. The scaled deflection 
angle $\vec{\alpha}$ is related to the true deflection angle $\hat{\alpha}$ by
$\vec{\alpha} = {{D_{od}D_{ds}}\over{{\xi_{0}}D_{os}}}\hat{\alpha}$, 
where $\xi_{0}$ is an arbitrary scale factor. 
If the lens is symmetric around the line of sight, equation 7 reduces to a function of only the radial
component $\vec{x}$,  and $\vec{\alpha}$ lies along the radial direction $\vec{x}$.

\begin{equation}
{\alpha(x) = {2\over{x}} {\int_0^{x}{x'\kappa(x')dx'}}}
\end{equation}
Thus, to determine the lensing contribution of the cluster, 
we must first model $\kappa(x)$.

From equation 3, we find that the three dimensional 
mass density at a radius r can be expressed as,

\begin{equation}
\rho_{grav}(r)={{3\beta kT}\over{4\pi \mu m_p G {r_c}^2}}{{(3+(r/r_c)^2)}\over{
(1+(r/r_c)^2)^2}}
\end{equation}
This quantity can be projected on the plane of the sky 
(normal to the line of sight),  resulting in the following 
expression for the surface mass density, $\Sigma(x)$,
as a function of the cylindrical radius $x$:

\begin{equation}
\Sigma(x)={{3kT\beta}\over{2 \mu m_p {\pi}G}}\,{{\left ({r_c^2\sqrt{(r_c^2+x^2)
(R^2-x^2)}+(2r_c^4+2r_c^2R^2+r_c^2x^2+x^2R^2)\Phi}\right)}\over{(r_c^2+x^2)^
{3/2}(r_c^2+R^2)}}
\end{equation}
where,$$\Phi = tan^{-1}{{\sqrt{{R^2-x^2}\over{r_c^2+x^2}}}} $$ 
and R is the radial extent of the cluster. 

Notice that the surface mass density of the cluster and the critical 
mass density both depend linearly upon $H_0$, thus implying that $\kappa$
does not depend on $H_0$. However, in our derivation of the 
convergence we infer the temperature of the cluster through
an empirical  luminosity - temperature  relation and therefore our results 
do depend on the assumed value for $H_{0}$.

In determining $\kappa(x)$ we have examined a variety of models in which we 
used a combination of different cluster core parameters.
We chose cluster limits ranging between  120$''$ and 280$''$.  We also varied the core 
radius between the values of $\theta_c = 5''$ and $\theta_c = 45''$ and 
use $H_{0}$ values of 50 and 75 km s$^{-1}$ Mpc$^{-1}$.
Note that at the smaller radii, the model approaches the limit of a 
singular isothermal sphere. We find that the average value of $\kappa$, at 
the location of the images, varies in the range of 0.07 to 0.21 as shown in Figure 6.
The upper and lower bounds on $\kappa$ place respective upper and lower 
bounds on $H_0$ in accordance with the models of Falco et al. (1991) and GN. 
For the estimation of $H_{0}$ we have taken a self consistent iterative 
approach. We begin with an initial trial value for $H_{0}$ and derive 
the total number of counts and cluster luminosity for an assumed core radius and cluster extent.  
Using the L$_{X}$ - T relation, we derive the cluster temperature and from 
equations 3 and 5  we obtain the total mass and convergence of the cluster, respectively.
The GN result (equation 4) provides an estimate for H$_{0}$. We reuse this 
value for H$_{0}$ to initialize the iterative procedure and repeat it
until the value for H$_{0}$ converges ($\Delta$H$_{0}$ $\leq$ 1). The result
is very insensitive to the initial guess for H$_{0}$. The error for 
H$_{0}$ is calculated by applying the self consistent iterative
approach for cluster core radii ranging between 5$''$ and 45$''$, 
cluster radii ranging between 120$''$ and 280$''$ and changing the 
mean HRI count rate by $\pm$ 2$\sigma$.
Assuming $q_0=0$, we obtain $ 67 <$ H$_{0}$ $<$ 82 $\,$ km $\,$ s$^{-1}$ $\,$ Mpc$^{-1}$ 
for a time delay of 1.1 years.

The effect of the cluster involves a rotation and an elongation of the separation vector 
connecting the two images. We may express the deflection angles
for images A and B as $\vec{\alpha}_{A} = \kappa_{A}{\vec{\theta}_{OA}}$ 
and  $\vec{\alpha}_{B} = \kappa_{B}{\vec{\theta}_{OB}}$,
respectively, where the subscript on $\kappa$ indicates the point of evaluation and where 
${\vec{\theta}_{OA}}$ and ${\vec{\theta}_{OB}}$ are the angular distances from the cluster
center to the image locations of A and B, respectively. 
The calculated values for the deflections $\vec{\alpha}_{A}$
and $\vec{\alpha}_{B}$ are shown in Figure 7. The fractional elongation of the angular separation
of images A and B produced by the cluster potential alone is 
${\left|\vec{AB}\right| - \left|\vec{A'B'}\right|}\over{\left|\vec{AB}\right|}$
and the calculated values for the percent elongation are presented in Figure 8;
here ${\vec{AB}}$ and ${\vec{A'B'}}$ denote, respectively, the vector separations
of the images with and without the cluster contribution.   

To calculate the convergences, $\kappa_{A}$ and  $\kappa_{B}$, we
use equation 8 with angular distances between
images A and B and the cluster center given by $x_{A}$ = 26.57$''$
and $x_{B}$ = 21.04$''$, respectively, based on the published position (-13.7$''$, -19.6$''$) 
(Angonin-Willaime et al. 1994) for the estimated cluster center of mass  
with respect to image B and the measured position of image A with respect 
to image B of (-1.25252$''$ ,6.04662$''$) (Gorenstein et al. 1984).  
Again, we use a grid of models in which we 
vary the cluster core radius from 5$''$ to 45$''$ and the cluster extent from 120$''$ to
280$''$.

\section{SPECTRAL ANALYSIS}
The quasar Q0957+561 also was observed with the {\it{ROSAT}} PSPC
on 1991 November 14 for 19006 sec, on 1993 November 8 for 4972 sec, 
and with the ASCA GIS on 1993 May 9 for 29000 sec. A summary of information
related to each of these observations is presented in Table 1. 
The spectral analysis for the 1991 November ROSAT observation was presented in
Chartas et al. (1995). 
The 1993 November PSPC observation was centered on Q0954+556 leaving the quasar
images located at an off-axis angle of about 47 arcmin. For the analysis we extracted 
the source counts within a circular region centered on Q0957+561 with a radius of 6 arcmin. 
Background counts were extracted from several other circles of 6 arcmin radius with centers located at 
an off-axis angle of 47 arcmin,  equal to that of Q0957+561.
To model the spectral response of the PSPC, we used the September 1994 PSPC detector 
response matrix pspcb\_gain2\_256.rmf provided by the ROSAT guest observer facility (GOF).
The spectrum was grouped to obtain a minimum of 20 counts in each energy bin,
allowing use of $ \chi^{2}$ statistics.

A simple redshifted power law plus galactic absorption due to cold material 
with solar abundances, provides an acceptable 
$\chi^{2}$ per degree of freedom (see Table 2). However, the resulting power law 
is slightly steeper (${\Delta}{\alpha}_{\nu} \sim 0.3$) than expected for a radio loud quasar 
at a redshift of 1.4 (Wilkes \& Elvis 1987; Shastri et al. 1993, Laor et al. 1997, Schartel et al. 1996).

In general the results obtained from the spectral fits to the 1993 November PSPC data 
are consistent with the results from the 1991 November PSPC observation of Q0956+561. 
In particular we find in both cases that an absorbed redshifted power law 
of photon index $\alpha_{\nu}$ plus a thermal Raymond-Smith model provides 
an acceptable $\chi^{2}$ fit to the data and yields 
values for the fitted parameters $\alpha_{\nu}$, $T_{e}$ and $N{_H}$ that are close to what 
is expected for a radio loud quasar at z=1.4. 
The X-ray flux of the thermal component 
derived from the fit to the PSPC spectrum of Q0957+561 is about 
5 $\times$ 10$^{-13}$ ergs s$^{-1}$ cm$^{-2}$, at least a factor of five greater 
than the X-ray flux of the cluster as
derived from the ROSAT HRI observations (see Table 3). This result implies that the 
observed ROSAT PSPC thermal component most likely originates 
from a source other than the cluster (see, e.g., Chartas et al. 1995).       

To constrain the model parameters even further we performed simultaneous fits to the 1991 November
and 1993 November PSPC data. We allowed the normalization of the power law and thermal 
component of each observation to vary independently. 

A comparison of the 0.2 - 2keV X-ray fluxes between the 1991 and 1993
ROSAT PSPC observations of Q0957+561 implies a decrease in flux over this interval by
$(25 \pm 9)\%$ with the quoted error being at the 90$\%$ confidence level.

Q0957+561 was also observed with the ASCA satellite during a pointed observation of NGC 3079 
on 1993 May 9 for 29072 sec.
In this observation  Q0957+561 is located 14.1$'$ off axis in the GIS2 image
and 9.28$'$ off axis in the GIS3 image
just within the field of view of the GIS2 and GIS3, but unfortunately outside the 
solid-state imaging spectrometer's (SIS's) field of view. 
The source counts were extracted from circles of radius 3.5 arcmin
centered on Q0957+561. The background was extracted from several other circles of the 
same radius centered at an off-axis angle equal to that of Q0957+561, but far away in azimuth.
For the data analysis of the GIS2 and GIS3 spectra we used the spectral response functions 
gis2v4\_0.rmf and gis3v4\_0.rmf, respectively, provided by the ASCA GOF.
The spectral data were grouped to obtain a minimum of 20 counts per bin. 
The observed ROSAT PSPC and ASCA GIS spectra are shown in Figure 9. 
The few source counts obtained in the ASCA observation prevent us from fitting
any complex models; however, a comparison of the spectral slope
of a simple power law model fitted to the 0.5-6 keV GIS data to that of the slope obtained
from fits to the 0.2-2 keV ROSAT data was possible. 
In particular a simultaneous fit of a simple absorbed power law
model between 0.5 and 5keV to the GIS2 and GIS3 spectra of Q0957+561
yields a photon index $\alpha_{\nu} = 1.6 \pm 0.1 $,
quite typical for radio loud quasars at z=1.4; a
simultaneous fit between 0.2 and 2keV of a similar model
to the November 1991 and November 1993 ROSAT PSPC spectra
gave a photon index of $2.27 \pm 0.03$. This difference may indicate 
a break in the spectrum at some energy above
2 keV. Alternatively, including a thermal component (fits 13 and 14) results 
in values of the photon index of the power law component
consistent with that for radio loud quasars.

\normalsize

\section{DISCUSSION and CONCLUSIONS}

The need for a detailed measurement 
of the mass distribution of the cluster lens has been emphasized by 
Falco et al.(1991) and by GN (1996) in analyses of 
lens models for Q0957+561 that incorporate the cluster contribution through 
a convergence parameter $\kappa(r)$.
The most important result of the present analysis of the ROSAT HRI data
of Q0957+561, that we hope will eventually lead to an accurate determination 
of the mass distribution of the cluster lens, is the 3$\sigma$ detection of X-rays from 
the cluster of galaxies that contains the principal lens galaxy G1. \\

As indicated in Figure 3, the uncertainty in 
our estimated values for the mass distribution of the cluster spans a large range
as do the consequent estimates of the convergence parameter of the lens (Figure 6).
The main contributors to this uncertainty are the lack of knowledge of the 
shape of the cluster, the core radius of the cluster, 
the temperature profile of the hot gas, the cluster extent, and the location of 
the cluster center as well as the sensitivity to the value of $H{_0}$ used for 
deriving the cluster hot gas temperature from the X-ray luminosity.
The present analysis yields a total cluster mass in the range 
1.5 - 3.2 $\times$ 10$^{14}$M${\odot}$, and does not include allowances
for errors in several of the above characteristics, but does consider
the effects of cluster core radii between 5$''$ and 45$''$,
and cluster extents between 120$''$ and 280$''$. The calculations 
of the X-ray luminosity were performed for Hubble constants of 50 and 
75 km s$^{-1}$ Mpc$^{-1}$.
The average convergence parameter $\overline{\kappa} = ({\kappa}_{A} + {\kappa_B})/2$ 
of the cluster at the image locations 
A and B was found to lie between 0.07 and 0.21. \\

One of the other assumptions in the present calculation
is that the temperature of the cluster follows the luminosity 
temperature relation of David et al. (1993) for a large sample of 
clusters of galaxies. The derived cluster temperature depends weakly on the cluster
X-ray luminosity. An error of a factor of 2 in the X-ray luminosity
derived from the HRI data will propagate as an error of about 
30$\%$ in the cluster temperature and about 7$\%$ (for $\kappa$ = 0.2) in the derived 
value of the Hubble constant.
The residual rms scatter in the $L_{X}$ - T relation along the temperature axis 
is $\sigma_{logT}$ = 0.104/2.1 according to Markevitch (1998) and propagates as
an error of about 3$\%$ in the derived value of the Hubble constant. 

 
Our calculations for the deflection angles produced from considering 
the influence of the cluster alone (not including G1) found the deflection for 
images A and B to lie in the range of 2$''$ to 16$''$ as shown in Figure 6;
the differences in these deflections for images A and B are much smaller. 
The resulting percent contribution, ${\Delta}S_{AB}$, to the observed 6.$''$1
separation of images A and B due to the cluster 
alone, lies in the range of 4\% to 22\%.
This range includes only the effects of varying the cluster core radius 
between 5$''$ and 45$''$ and the cluster limit between 120$''$ and 280$''$.
This result is in contrast with the commonly accepted notion that the large separation
of the two images of Q0957+561 is an indicator that the lens
contains a cluster of galaxies. To test the sensitivity of this result to 
the adopted cluster center position, we computed the cluster contribution to the 
image separation for cluster center locations within a distance of 5$''$
of the location derived from the optical observations by
Angonin-Willaime et al. (1994).
As expected, when the cluster center is moved closer to the locations
of images A and B, the derived cluster contribution to the image separation becomes 
more significant reaching a value as high as ${\Delta}S_{AB}$ = 31\% for a cluster center 
at a distance of about 16$''$ from image B. \\


Recently the mass distribution of the lens of Q0957+561 was determined
by Fischer et al. (1997) based on the distortion of background galaxies produced 
by the lens. Their estimate of 3.7 $\pm$ 1.2 $\times$ 10$^{14}$M${\odot}$ 
within a radius of 1Mpc is consistent with the X-ray estimates.
As pointed out by Schneider \& Seitz (1995) however, the Kaiser \& Squires (1993)
mass reconstruction technique also is insensitive to constant density sheets of matter.


An interesting conclusion from the mass reconstruction analysis by Fischer et al. (1997) 
is the relatively low value of 5$''$ derived for the cluster core radius.
The cluster core radius, r$_{cF}$, as defined in Fischer et al. (1997), however, differs 
from the cluster core radius, r$_{c}$, used in this paper and defined in equation 2.
The relation between r$_{c}$ and r$_{cF}$ is given approximately by r$_{cF}$ = -0.987 + 0.685r$_{c}$
for 5$''$ $<$ r$_{c}$ $<$ 50$''$. We will have to await future X-ray 
observations of Q0957+561 to compare more accurately  X-ray and weak-lens 
results for the cluster core radius. \\

The present ROSAT detection of cluster emission suggests that future X-ray 
observations will greatly add to our understanding of Q0957+561.
AXAF observations should provide direct determination of the temperature profile,
the core radius and the shape of the cluster lens, reveal possible clumpiness
in the total mass profile, and ultimately
provide a tighter constraint on the Hubble constant.\\

\acknowledgements

We are grateful to Emilio Falco for providing many useful comments 
and to Donald Schneider for making many useful suggestions on
several early drafts of this paper. This research was supported by NASA through
grant NAS 8-38252.

\appendix

\newpage

\beginrefer

\refer Angonin-Willaime M.-C., Soucail, G., \& Vanderriest C., 1994, A\&A, 291, 411. 

\refer Arnaud, K., Johnstone, R., Fabian, A., Crawford, C., Nulsen, P., Shafer, R., 
and Mushotzky, R. 1987, MNRAS., 227, 241.

\refer Bar-Kana, R., 1997, electronic preprint astro-ph/9701068.

\refer Bernstein, G., Fischer, P., Tyson, J. A., and Rhee, G., 1997, ApJ, 483, L79.

\refer Chartas, G., Falco, E. E., Forman, W. R., Jones, C., Schild, R., \&  Shapiro, I. I.,
 1995, ApJ, 445, 140.

\refer Corbett, E. A., Browne, I. W. A., Wilkinson, P. N. \& Patnaik, A. R. 1996,
in Astrophysical Applications of Gravitational Lensing, ed. Kochanek, C. S., \&  Hewitt, J. N.,
Kluwer: Dordrecht, p. 37.

\refer David, L. et al., 1993, ApJ, 412, 479. 


\refer Falco, E. E., Gorenstein, M., \& Shapiro, I., 1991, ApJ, 372, 364.

\refer Falco, E. E., Shapiro, I. I., Moustakas, L. A., \& Davis, M., 1997, ApJ,
484, 70.

\refer Fischer, P., Bernstein, G., Rhee, G. \& Tyson, J. A. 1996, AJ, 113, 521.

\refer Geiger, B., \& Schneider, P., 1996, MNRAS, 282, 2, 530.

\refer Gorenstein, M. et al., 1984, ApJ, 287, 538.

\refer Grogin, N., \& Narayan, R., 1996, ApJ, 464, 92.

\refer Grogin, N., \& Narayan, R., 1996, erratum ApJ, 473, 570.


\refer Haarsma, D. B., Hewitt, J. N., Leh\'{a}r, \& Burke, B. F., 1997,
ApJ, 479, 102.

\refer Henriksen, M. 1985 Ph.D. thesis, University of Maryland.

\refer Hughes, J., Yamashita, K., Okumura, Y., Tsunemi, H., and Matsuoka, M., 1988, ApJ, 327, 615.

\refer Jones, C., Forman, W., Stern, C., Falco, E., \& Shapiro, I.
1993, ApJ, 410, 21.

\refer Keeton, C. R., Kochanek, C. S., 1997, ApJ, 487, 42.

\refer Kundi\'{c}, T. et al. 1997, ApJ, 482, 75.

\refer Laor, A., et al. 1997, ApJ, 477, 93L. 

\refer Markevitch, M., 1998, ApJ, submitted (astro-ph 9802059)

\refer Pelt, J., Kayser., R., Refsdal., S., Schramm, T., 1996, A\&A, 305 97.

\refer Refsdal, S., 1964, MNRAS, 128, 295.

\refer Refsdal, S., 1964, MNRAS, 128, 307.

\refer Refsdal, S., 1966, MNRAS, 132, 101.

\refer Roberts, D. H., Lehar, J., Hewitt, J. N. \& Burke, B. F., 1991, Nature, 352, 43.

\refer Schartel, N., et al. 1996, MNRAS, 283, 1015.

\refer Schechter, P. L. et al., 1997, ApJ, 475, L85.

\refer Schneider, P., Ehlers, J., \& Falco, E. E., 1992, Gravitational Lensing (New York: Springer)

\refer Schild, R. \& Thomson, D. J., 1995, AJ., 109, 1970.

\refer Stark, A. A., Gammie, C. F., Wilson, R. W., Bally, J., Linke, R. A., Heiles, C., 
\& Hurwitz, M., 1992, ApJS, 79, 77.

\refer Vanderriest, C., Schneider, J., Herpe, G., Chevreton, M., Moles, M. \& Wlerick, G., 1989, A\&A, 215, 1.

\refer Walsh, D., Carswell, R. F., \& Weymann, R. J., 1979, Nature, 279, 381.

\endrefer

\clearpage


\figcaption[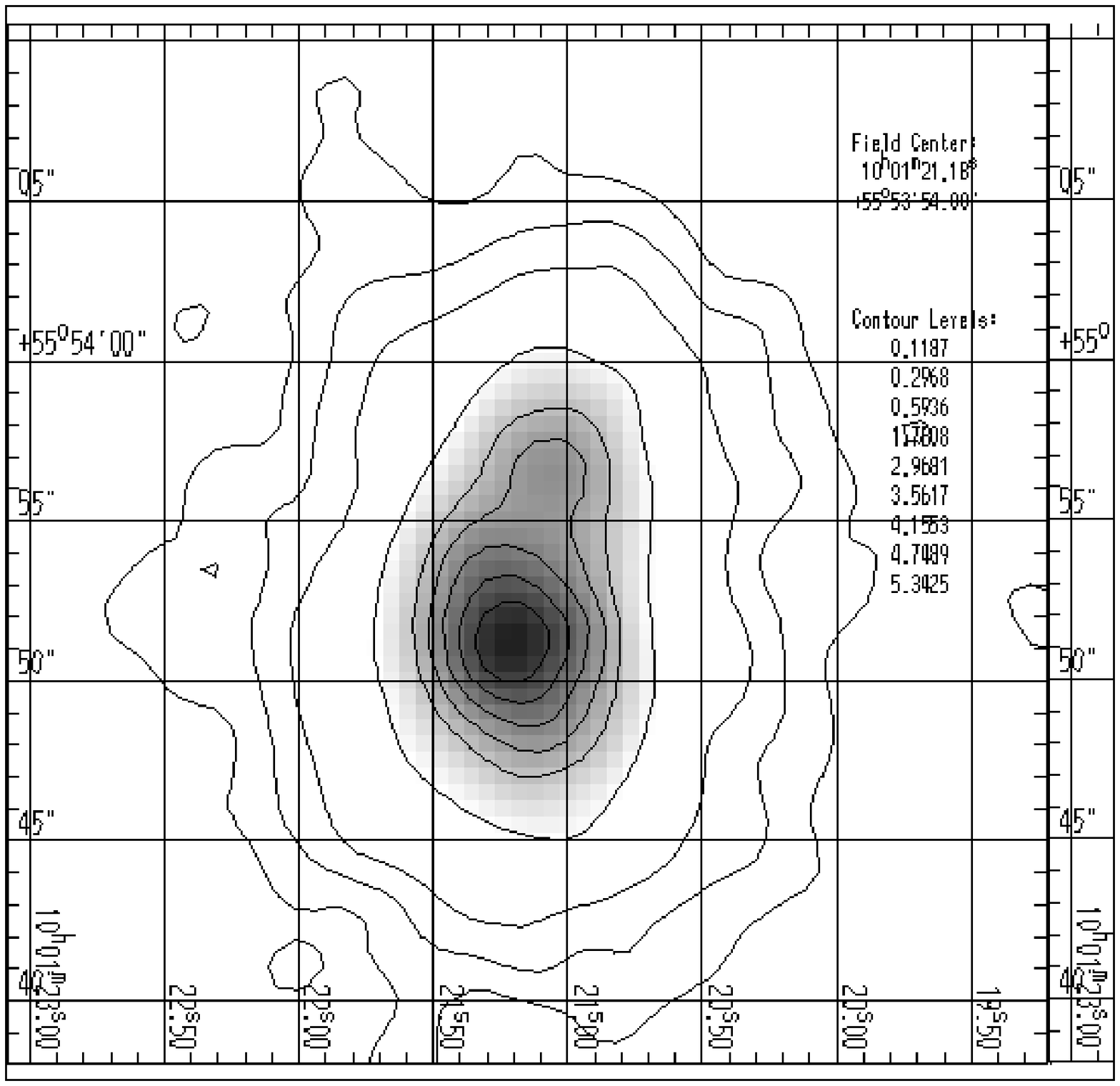]{Composite iso-intensity plot of three co-aligned ROSAT HRI observations of Q0957+561.
Contour levels are expressed as a percentage of the peak value.}

\figcaption[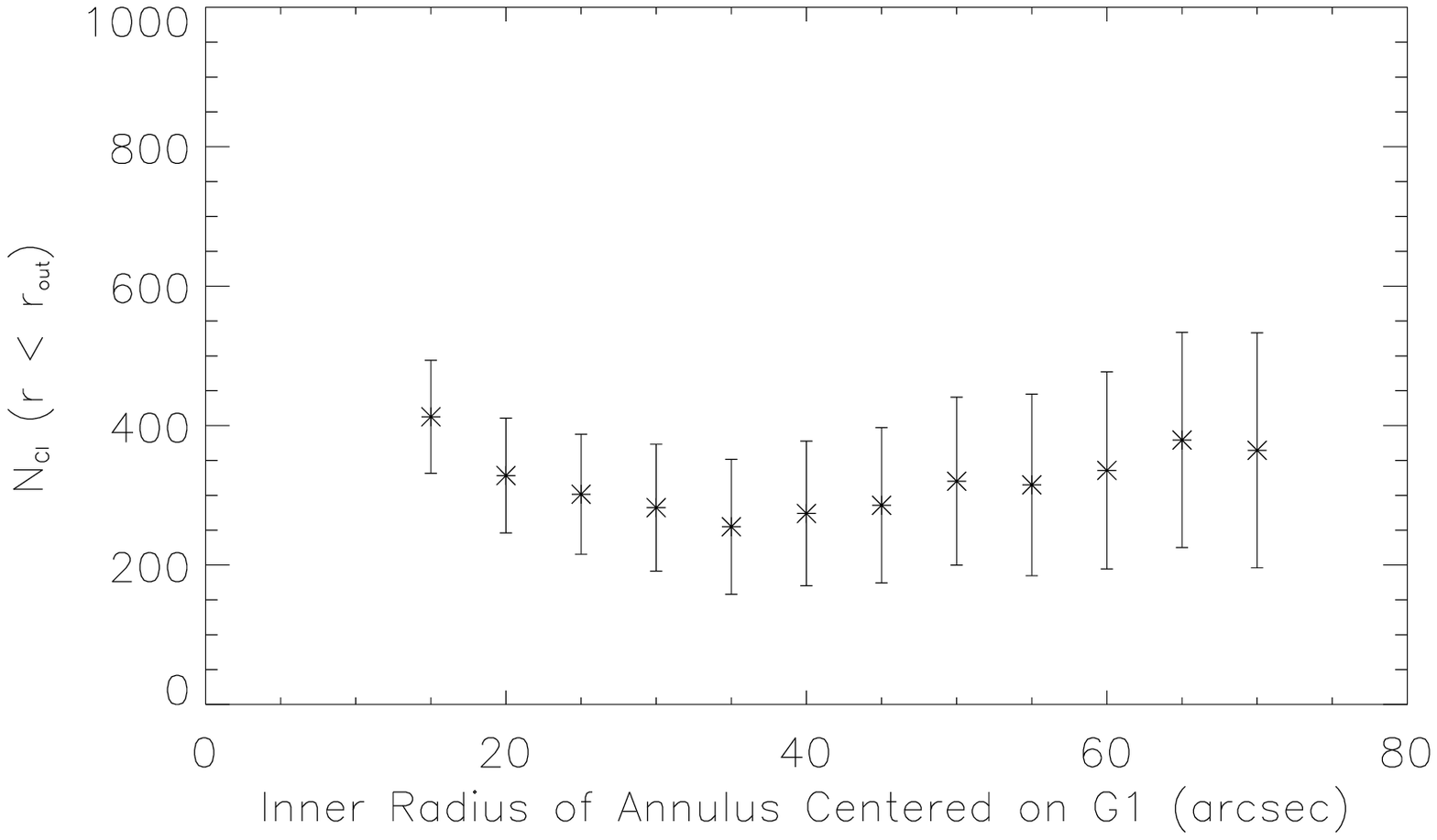]{The total counts from the ROSAT HRI, $N_{Cl}$, within a 
circular region centered on Q0957+561 and originating from cluster emission. This number 
was estimated by extracting from the HRI image of the Q0957+561 system the number of 
counts within a selected annulus centered on Q0957+561 and applying a correction for 
the expected counts residing outside this annulus. For this plot, we used a 
cluster core radius of 45$''$ and a cluster extent of 280$''$.}

\figcaption[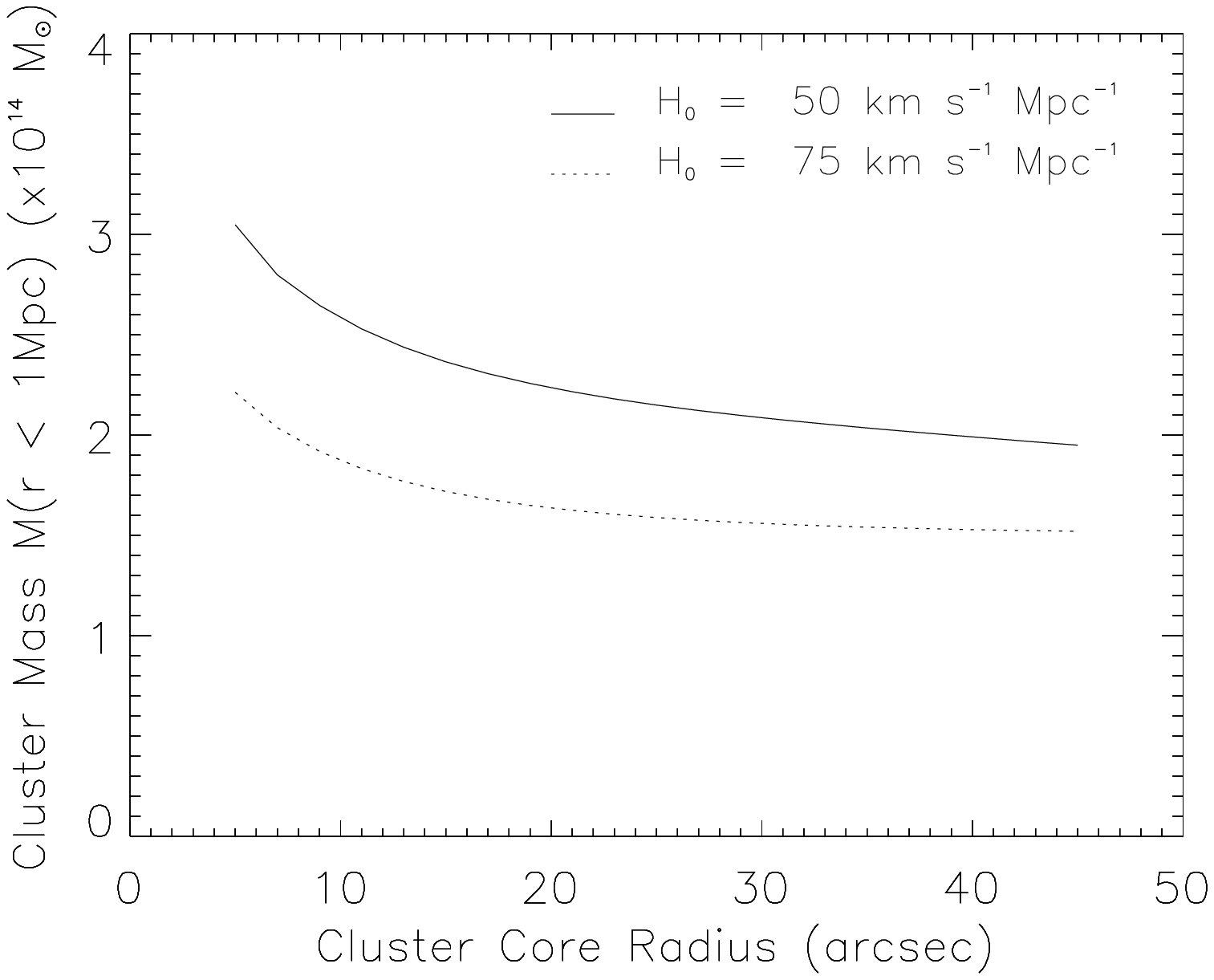]{Estimated total mass of the cluster containing the G1 galaxy within a radius of 1Mpc
over a range of typical cluster core radii, for q$_{0}$ = 0 and for $H_{0}$ values of 50 and
75 km s$^{-1}$ Mpc$^{-1}$.}

\figcaption[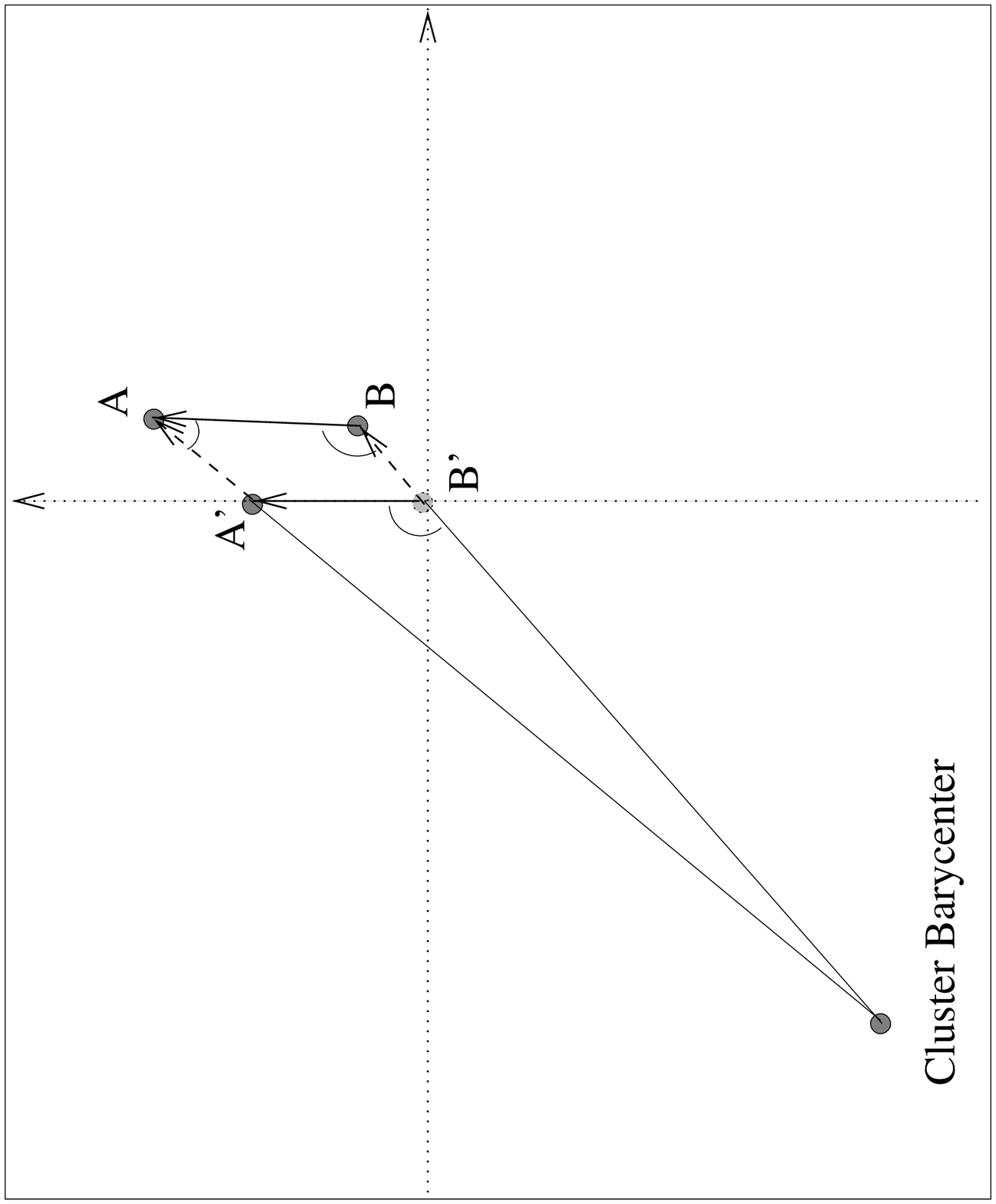]{Conceptual diagram of the cluster's contribution 
to the deflection effect. Images A and B are the observed lensed images 
of Q0957+561 while images A$'$ and B$'$ are the expected images for  
deflection produced from the G1 galaxy alone.}

\figcaption[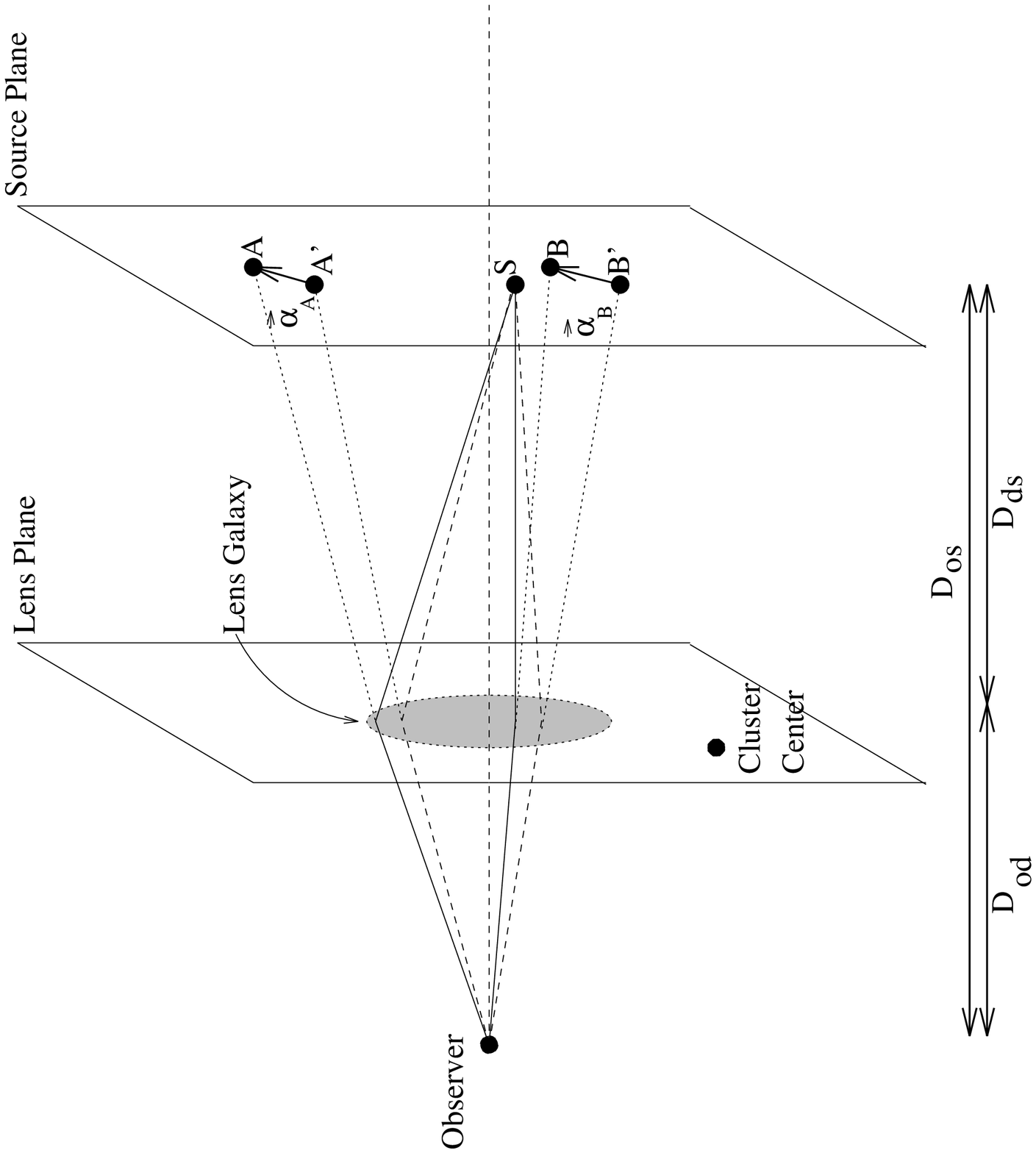]{Basic lens geometry of the 
system Q0957+561. The shown deflection angles $\vec{\alpha}_{A}$
and $\vec{\alpha}_{B}$ are due to the cluster contribution alone.}

\figcaption[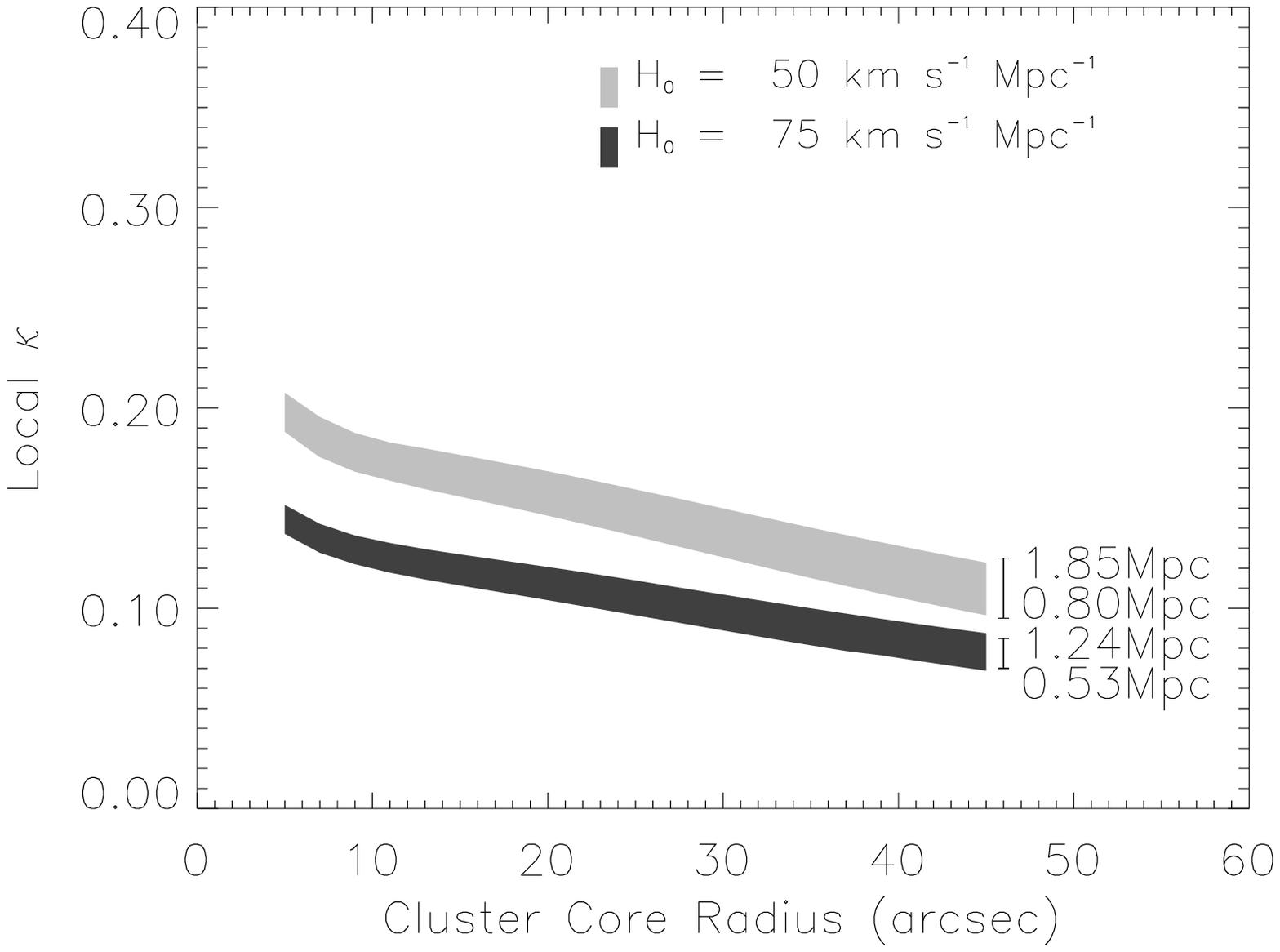]{The local convergence parameter $\kappa$, averaged over locations of images
A and B, as estimated for cluster core radii ranging  between 5$''$ and 45$''$,
and cluster limits ranging between 120$''$ and 280$''$. The calculations 
of the X-ray luminosity were performed for Hubble constants of 50 and 
75 km s$^{-1}$ Mpc$^{-1}$ and for q$_{0}$ = 0. This range of uncertainty in the convergence 
parameter $\kappa$  does not include possible systematic errors arising from 
the derivation of the cluster temperature through the use of the 
luminosity-temperature relation for clusters of galaxies or from
other assumptions, such as the shape of the cluster.}

\figcaption[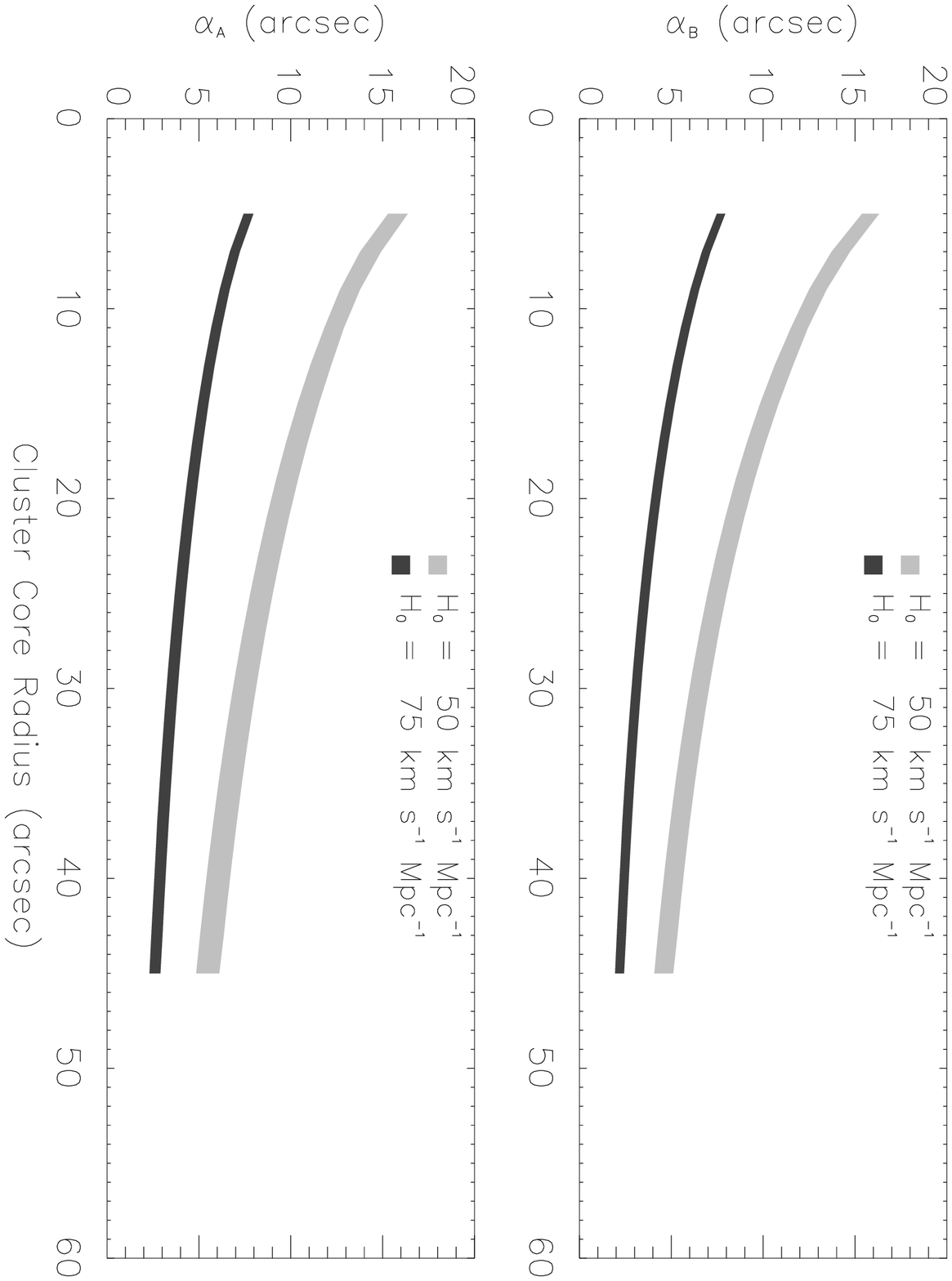]{Calculated values for the deflection angles $\vec{\alpha}_{A}$
and $\vec{\alpha}_{B}$ produced by the cluster. The cluster core radius,
the cluster extent, and $H_{0}$ were varied within the ranges specified in 
the caption for Figure 6. For other relevant assumptions see text.}

\figcaption[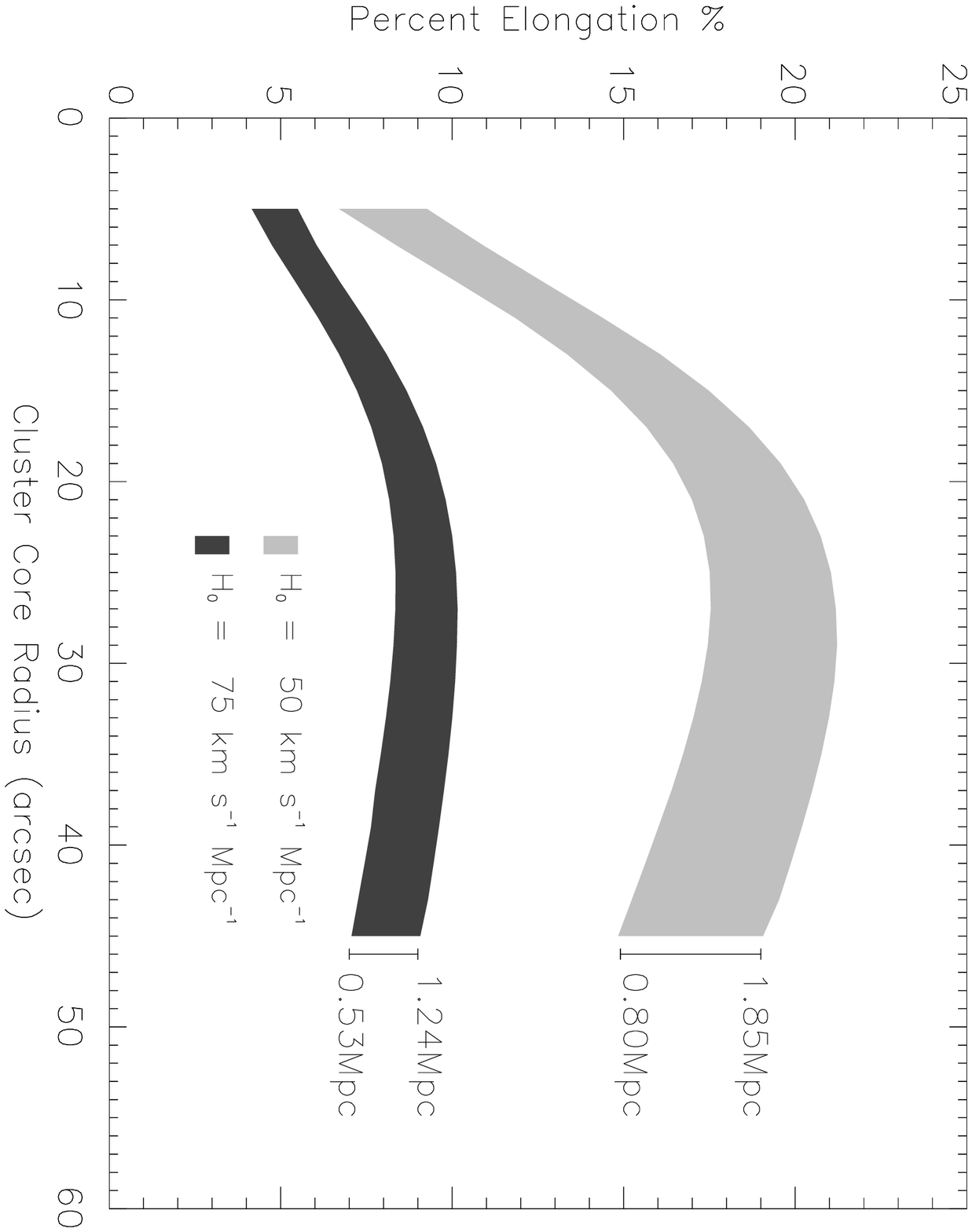]{The percent elongation of the angular separation
of images A and B produced by the cluster potential alone. The cluster core radius,
the cluster extent and $H_{0}$ were varied within the ranges specified in the caption
for Figure 6. For other relevant assumptions see text.}

\figcaption[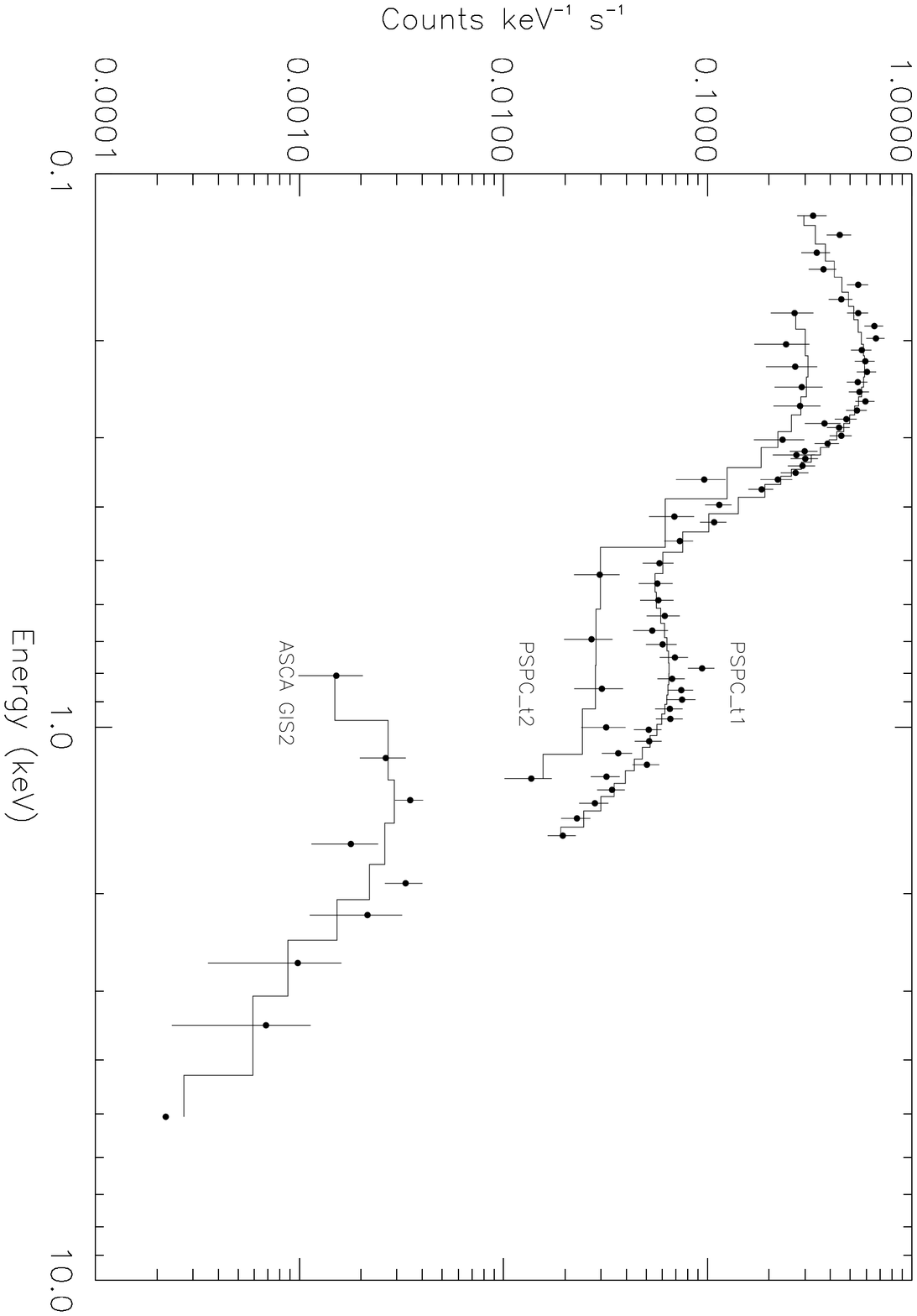]{Simultaneous spectral fit to ROSAT PSPC data taken on 1991 November 15
and 1993 November 8, and to ASCA GIS data taken on 1993 May 9. The model used
for this fit incorporates a redshifted power law plus a thermal Raymond-Smith
model and galactic absorption (see text).}



\clearpage

\scriptsize
\begin{center}
\begin{tabular}{llllll}
\multicolumn{6}{c}{TABLE 1} \\
\multicolumn{6}{c}{{\it{ROSAT}} HRI, PSPC and ASCA GIS Observations of Q0957+561} \\
& & & & &   \\ \hline\hline
\multicolumn{1}{c} {Observation} &
\multicolumn{1}{c} {Instrument} &
\multicolumn{1}{c} {Observation} &
\multicolumn{1}{c} {Exposure} &
\multicolumn{1}{c} {Off axis} &
\multicolumn{1}{c} {Net Counts} \\

{Date}           &                   & File name Id      &       &angle      & \\
                 &                   &                   &$sec$  &$arcsec$   & \\ \hline
1991 May 1       & {\it{ROSAT}} HRI  & rh700100          & 20847 &   45      &  473 $\pm$ 30 \\   
1992 October 20  & {\it{ROSAT}} HRI  & wh600411n00       & 20676 &  740      &  812 $\pm$ 34 \\
1992 October 22  & {\it{ROSAT}} HRI  & rh700889n00       & 18548 &   50      &  800 $\pm$ 33 \\
1994 April 11    & {\it{ROSAT}} HRI  & rh701294n00       & 25083 &   50      & 1077 $\pm$ 39 \\
1991 November 15 & {\it{ROSAT}} PSPC & rp300319          & 19006 &  993      & 3677 $\pm$ 70 \\
1993 November 8  & {\it{ROSAT}} PSPC & rp701401n00       &  4472 & 2821      &  449 $\pm$ 27 \\
1993 May 9       & {\it{ASCA}} GIS2  & ad60000000g200470 & 29072 &  846      &  140 $\pm$ 20 \\
1993 May 9       & {\it{ASCA}} GIS3  & ad60000000g300470 & 29072 &  557      &  121 $\pm$ 18 \\
\hline \hline
\end{tabular}
\end{center}

\noindent 
NOTE- A description of the format for the ROSAT high resolution instrument (HRI),
ROSAT position sensitive proportional counter (PSPC) and ASCA gas imaging spectrometer (GIS)
file names is available at the ROSAT and ASCA guest observer facility web pages. 

\newpage

\scriptsize
\begin{center}
\begin{tabular}{lllllll}
\multicolumn{7}{c}{TABLE 2} \\
\multicolumn{7}{c}{Model Parameters Determined from Spectral Fits to the} \\
\multicolumn{7}{c}{{\it{ROSAT}} PSPC and ASCA GIS Spectra of Q0957-561} \\
& & & & & & \\ \hline\hline
\multicolumn{1}{c} {Fit} &
\multicolumn{1}{c} {Model$^{a}$} &
\multicolumn{1}{c} {Instruments$^{b}$} &
\multicolumn{1}{c} {$\alpha_\nu$} &
\multicolumn{1}{c} {$N_{H}(z=0)$} &
\multicolumn{1}{c} {$kT_{e}(keV)$} &
\multicolumn{1}{c} {$\chi^{2}/dof(dof)$} \\
 &   &                         &                          &{$10^{20}$$cm^{-2}$}&                        & \\ \hline
1& 1 &$PSPC_{t2}$              & $2.36^{+0.11}_{-0.10}$   & 0.87                     &                        & 0.95(13)  \\
2& 2 &$PSPC_{t2}$              & $2.97^{+0.45}_{-0.40}$   & $2.16^{+0.99}_{-0.84}$   &                        & 0.80(12)  \\
3& 3 &$PSPC_{t2}$              & $1.71^{+0.44}_{-0.78}$   & 0.87                     & $0.43^{+0.11}_{-0.18}$ & 0.73(11)  \\
4& 1 &$PSPC_{t2} + PSPC_{t1}$  & $2.27^{+0.028}_{-0.027}$ & 0.87                     &                        & 0.88(63) \\
5& 2 &$PSPC_{t2} + PSPC_{t1}$  & $2.55^{+0.11}_{-0.08}$   & $1.44^{+0.22}_{-0.16}$   &                        & 0.69(62) \\
6& 3 &$PSPC_{t2} + PSPC_{t1}$  & $2.06^{+0.07}_{-0.08}$   & 0.87                     & $0.45^{+0.08}_{-0.08}$ & 0.62(60) \\
7& 4 &$PSPC_{t2} + PSPC_{t1}$  & $2.14^{+0.85}_{-0.19}$   & $0.99^{+0.44}_{-0.25}$   & $0.42^{+0.09}_{-0.28}$ & 0.63(59) \\
8& 1 & GIS2                    & $1.29^{+0.12}_{-0.022}$  & 0.87                     &                        & 1.85(12) \\
9& 1 & GIS3                    & $1.91^{+0.19}_{-0.17}$   & 0.87                     &                        & 1.34(7)  \\
10&1 & GIS2 + GIS3             & $1.58^{+0.12}_{-0.12}$   & 0.87                     &                        & 1.99(20) \\
11&1 &$PSPC_{t2} + PSPC_{t1}$ + &&&& \\
&    & GIS2 + GIS3             & $2.25^{+0.07}_{-0.08}$   & 0.87                     &                        & 1.42(84) \\
12&2 &$PSPC_{t2} + PSPC_{t1}$ + &&&& \\
&  & GIS2 + GIS3               & $2.33^{+0.07}_{-0.08}$   &$1.04^{+0.15}_{-0.14}$    &                        & 1.42(83) \\
13& 3 &$PSPC_{t2} + PSPC_{t1}$ + &&&& \\
&   & GIS2 + GIS3                & $1.89^{+0.08}_{-0.08}$   & 0.87                   & $0.41^{+0.05}_{-0.05}$ & 1.09(79) \\
14& 4 &$PSPC_{t2} + PSPC_{t1}$ + &&&& \\
&  & GIS2 + GIS3               & $1.58^{+0.14}_{-0.14}$   & $0.42^{+0.14}_{-0.14}$ & $0.5 ^{+0.1}_{-0.1}$   & 1.01(78) \\
\hline \hline
\end{tabular}
\end{center}

\noindent 
NOTES-\\
$ ^{a}$ Model 1 incorporates a redshifted power law plus absorption due to cold material at
solar abundances fixed to the galactic value. Model 2 incorporates a redshifted power law plus 
absorption due to cold material at solar abundances set as a free parameter in the fit.
Model 3 incorporates a redshifted power law plus thermal Raymond-Smith model plus
galactic absorption.
Model 4 incorporates a redshifted power law plus thermal Raymond-Smith model plus
absorption due to cold material at solar abundances set as a free
parameter in the fit.\\
$ ^{b}$ $PSPC_{t1}$ and $PSPC_{t2}$ correspond to the ROSAT PSPC observations made on 1991 November 15
and on 1993 November 8, respectively.\\

\newpage

\scriptsize
\begin{center}
\begin{tabular}{llllll}
\multicolumn{6}{c}{TABLE 3} \\
\multicolumn{6}{c}{X-ray Fluxes and Luminosities of the Thermal and Power Law Components} \\
& & & & &   \\ \hline\hline
\multicolumn{1}{c} {Fit} &
\multicolumn{1}{c} {Instrument}$$ &
\multicolumn{1}{c} {$f_{X,Thermal}^{a}$} &
\multicolumn{1}{c} {$f_{X,Power}^{a}$} &
\multicolumn{1}{c} {$L_{X,Thermal}^{b}$} &
\multicolumn{1}{c} {$L_{X,Power}^{b}$}  \\
&&$10^{-12}$ erg cm$^{-2}$ s$^{-1}$&$10^{-12}$ erg cm$^{-2}$ s$^{-1}$&$10^{46}$ erg s$^{-1}$&$10^{46}$ erg s$^{-1}$ \\ \hline
6 &$PSPC_{t1}$& $0.52 \pm 0.3$  & $1.03 \pm 0.19$      & 1.29      & 1.68      \\
6 &$PSPC_{t2}$& $0.57 \pm 0.3$  & $0.68 \pm 0.19$      & 1.42      & 1.11      \\
\hline \hline
\end{tabular}
\end{center}
\noindent
NOTES-\\
$ ^{a}$ Absorbed flux between 0.2 - 2 keV.\\
$ ^{b}$ Luminosity between 0.2 - 2 keV. No corrections due to magnification from 
lensing have been made. \\

\end{document}